\documentclass[floatfix,12pt]{iopart}
\usepackage{iopams}
\usepackage{graphicx}
\usepackage{subfigure}
\usepackage{psfrag}
\usepackage{stackengine}
\usepackage{color}

\begin{document}

\title{Fluctuations in interacting particle systems with memory}
\author{Rosemary J\ Harris}
\address{School of Mathematical Sciences, Queen Mary University of London, Mile End Road, London, E1 4NS, UK}
\ead{\mailto{rosemary.harris@qmul.ac.uk}}

\begin{abstract}
We consider the effects of long-range temporal correlations in many-particle systems, focusing particularly on fluctuations about the typical behaviour.  For a specific class of memory dependence we discuss the modification of the large deviation principle describing the
probability of rare currents and show how superdiffusive behaviour can emerge.  We illustrate the general framework with detailed calculations for a memory-dependent version of the totally asymmetric simple exclusion process as well as indicating connections to other recent work.

\end{abstract}

\section{Introduction}
\label{s:intro}

Interacting particle systems in driven steady states are typically characterized by non-zero currents; among the recent advances in non-equilibrium statistical mechanics has been a considerable body of work on understanding the fluctuations of such currents.  Indeed it is now well-established that, for \emph{Markovian} dynamics, the probability of seeing a time-averaged current away from the mean is generically captured by a large deviation principle with ``speed'' $t$~\cite{Touchette09b,Derrida07b,Bertini06b}.  However, models with some form of \emph{non-Markovian} dynamics arguably describe better the long-range temporal correlations in many real scenarios~\cite{Mantegna99,Rangarajan03,Hoefling13}.  In this direction, there is topical interest in both the typical behaviour and fluctuations for particle systems with memory.  In particular, statistical physicists have recently studied a variety of memory-dependent random walkers in classical and quantum contexts, see e.g.,~\cite{Schutz04,Cressoni07,Serva13b,Rohde13} 
-- some of these can be related to the reinforced random walks and P\'olya urn models found in earlier mathematical literature and reviewed, for instance, in~\cite{Pemantle07}. 
Much less is known about non-Markovian \emph{many}-particle systems but some aspects of the stationary-state properties (e.g., mean current as a function of density, conditions for a condensation transition) have been investigated for models with internal states or non-exponential waiting times~\cite{Hirschberg09,Concannon14,Khoromskaia14}.   Going beyond the typical behaviour, the current fluctuations in a temporally-correlated zero-range process 
have also recently been explored (and compared to the equivalent memoryless model) although exact analytical calculations proved possible only for a single site~\cite{Cavallaro15}.

In the present contribution we build on earlier work in~\cite{Me09} to show how an expansion about fixed points of the dynamics can yield valuable information about the fluctuations in a particular class of non-Markovian interacting particle systems, even when full solution appears a formidable task.  Specifically, this enables us to predict the speed of the current large deviation principle and hence the long-time scaling behaviour of fluctuations.  We demonstrate this approach with perhaps one of the most famous models in non-equilibrium statistical mechanics: the totally asymmetric simple exclusion process (TASEP).  Here we show how a current-dependent input rate leads to a modified phase diagram including a superdiffusive regime and we check our theoretical approximations against simulations and exact numerics.

The remainder of the paper is structured as follows.  In section~\ref{s:frame} we introduce the framework of systems with current-dependent rates and indicate the connections to other recent works, including some of those mentioned above.  In section~\ref{s:fixed} we perform a stability analysis of fixed points and make a Gaussian expansion to study the fluctuations.  The power of this approach is then illustrated by treatment of the TASEP in section~\ref{s:ASEP} before a concluding discussion and wider perspective in section~\ref{s:dis}.  Finally, a short appendix provides a pedagogical treatment of a single-particle problem in order to demonstrate the formalism.  

\section{Interacting particle systems with current-dependent rates}
\label{s:frame}

We work within a discrete-space and continuous-time framework with the particle configuration at time $t$ labelled by $\sigma(t)$ and transition rate from state $\sigma$ to $\sigma'$ given by the matrix element $w_{\sigma',\sigma}$.   Classical lattice-based many-particle models described in this way include exclusion processes (to which we will return later)~\cite{Derrida98c,Spitzer70}, zero-range models~\cite{Spitzer70,Evans05}, and inclusion processes~\cite{Giardina07}.  

In systems of this type, a time-integrated particle current $\mathcal{J}(t)$ can be defined as the net number of jumps across a given bond (or subset of bonds) from time zero up to time $t$.  We choose the script style to indicate that the current is a functional of the stochastic history $\{\sigma(\tau),0\leq\tau\leq t\}$ but will suppress the explicit dependence on $t$ where no confusion should arise.  It is well-known that $\mathcal{J}$ generically obeys a large deviation principle which can be loosely stated as
\begin{equation}
\mathrm{Prob}\left(\frac{\mathcal{J}}{t}=j\right) \sim \rme^{-I_w(j) t} \label{e:ldp}
\end{equation}
where $\sim$ denotes logarithmic equivalence in the long-time limit. $I_w(j)$ is known as the rate function and $t$, somewhat misleadingly, referred to as the speed.  The $w$ subscript emphasizes that the rate function depends in some non-trivial way on the set of transition rates.  Much recent industry has been devoted to calculating $I_w(j)$ for various models both within a ``microscopic'' lattice-based approach~\cite{Derrida07b} and in the ``macroscopic'' hydrodynamic limit~\cite{Bertini06b}.

Here, following~\cite{Me09}, we introduce an element of memory by considering a class of models in which the rates at time $t$ depend on the current up to time $t$.   To be precise, the rates $w_{\sigma',\sigma}$ now depend on the time average $\mathcal{J}/t$ of some specified particle current and will be denoted by $w_{\sigma',\sigma}(j)$.  Obviously, the functional dependence on the current must be chosen so that the rates always remain positive.  To avoid singularities at time zero we also assume that the time-averaged current starts at some fixed value $j_0$ at a time $t_0$ which is small compared to the overall measurement time $t$.  Physically, taking the condition $0 \ll t_0 \ll t$ excludes any initial transient behaviour which could be governed by different rates.   A conceptually simple generalization is to the case where rates depend on multiple currents (e.g., currents measured separately across different bonds or in different directions) -- much of the following analysis can be extended to that situation although we shall not consider it in detail.   We emphasize that we specialize here to models with a functional dependence on the single variable $\mathcal{J}/t$ (\emph{time-averaged} current), rather than a more general dependence on $\mathcal{J}$ and $t$ separately.  The closely related scenario of feedback depending on the \emph{time-integrated} current has also recently been explored, for example, in a quantum context~\cite{Brandes10}. 

For illustrative purposes we now consider a single particle, i.e., a random walker, with this type of memory and endeavour to describe its connection to a range of models in the literature which may, at first sight, appear rather disparate.    The natural ``current'' for a particle on a one-dimensional lattice is just the net number of steps made in one direction, say towards the right, and we now assume left and right hopping rates which depend on the time average of this quantity (in other words, on the particle velocity).   
In the discrete-time version of this picture, the dependence is thus on the particle's position 
divided by the number of time steps elapsed.   Dynamics in this category includes the ``elephant'' random walker of~\cite{Schutz04}
as well as several other recent random walk scenarios~\cite{Hod04,Huillet08b,Kumar10c}, the voting model of~\cite{Hisakado10}, and some aspects of the behaviour
in a discrete-choice model with dependence on the peak of past experience~\cite{Me15}.   In fact, mathematically, these are all essentially equivalent to the much older P\'olya urn problem~\cite{Polya31} in which the probability for picking a black or white ball depends on the relative number (fraction) of such balls chosen in the past.  If the functional form of the dependence is non-linear, then one has a generalized P\'olya process, for overviews see, e.g.,~\cite{Pemantle07,Hill80}.  In passing, we remark that such models can also be considered as a limiting case of binary Markov chains with memory of a finite number of steps; see e.g.,~\cite{Usatenko03,Hisakado15} and note that the latter reference illustrates further connections to Kirman's ant colony model~\cite{Kirman93} (which has potential relevance to economic markets) and even the kinetic Ising model~\cite{Kawasaki72}.

Our focus here is on continuous-time models with dependence on the current over the whole history.  In the single-particle case we note that even when the particle remains stationary, the time-averaged current changes due to the continuous increase of time $t$ in the denominator of $\mathcal{J}/t$.  The dynamics of the particle can be thought of as a type of continuous-time random walk (CTRW) or ``semi-Markov'' process with a complicated non-exponential distribution of waiting times which, in general, also depends on the time of the last jump (so that successive waiting times are not identically distributed).  This correspondence is particularly clear in the case of a random walker moving only in one direction (see~\ref{A:single}) and provides a possible route to genuine continuous-time numerical simulations rather than the brute-force approach of using a discrete-time update rule with very short time steps.  However, the situation is more complicated for many-particle systems, or those with a dependence on multiple currents, and it may be practically difficult to obtain explicit forms for the relevant waiting time distributions.   There is a vast body of work on CTRWs with identically distributed, typically power law, waiting times (see~\cite{Burioni14b,Krusemann15} for just a couple of recent examples, discussing different scaling regimes and the effects of bias) as well as on more general time-homogeneous semi-Markov processes~\cite{Maes09b} and applications~\cite{Gorissen12c}.  Helpful explanations of the connections between different commonly-employed formulations can be found in~\cite{Goychuk04} and~\cite{Qian06}.

Feedback based on the time-averaged current clearly has the potential to introduce long-range temporal correlations and one might ask how these modify the current large deviation principle~\eref{e:ldp}, if indeed such a relationship still exists.    The chief result of~\cite{Me09} was that if, for some $\gamma$, the limit
\begin{equation}
\tilde{I}(j) = \lim_{t \to \infty} \min_{q(\tau)} \frac{1}{t^\gamma} \int_{t_0}^t {I}_{w(q)}(q+\tau q') \, d\tau \label{e:mldp}
\end{equation}
exists (and is not everywhere zero), then it is the rate function for a modified large deviation principle with speed $t^\gamma$.  In other words, we now have
\begin{equation}
\mathrm{Prob}\left(\frac{\mathcal{J}}{t} =j\right)\sim \rme^{-\tilde{I}(j) t^\gamma}, \label{e:mldp2}
\end{equation}
where $\gamma$ is not necessarily equal to unity.  Note that, in~\eref{e:mldp}, $I_{w(q)}$ is the \emph{Markovian} rate function evaluated with transition rates $w(q)$, and $q(\tau)$ is a trajectory in the space of time-averaged currents with fixed initial condition ($q(t_0)=j_0$) and final condition ($q(t)=j$).

This result can be derived heuristically by what has been dubbed a ``temporal additivity principle'' (a time-based analogue of the spatial additivity principle of Bodineau and Derrida~\cite{Bodineau04}) in which one notes that the time-averaged current changes very slowly for large times so can be approximated as constant over time slices long compared with the dynamics.  Carefully taking the limit $t \to \infty$ such that both the length and the number of the time slices becomes infinite,  this quasistatic (or adiabatic) argument gives an integral form for the probability of seeing a given path in current space.  Furthermore, in the long-time limit a particular current fluctuation is overwhelmingly likely to be realised by the optimal (or typical) path which is found by minimizing over all $q(\tau)$ consistent with the required initial and final current conditions.  For further details of this analysis we refer the interested reader to~\cite{Me09}.\footnote{Note that the technical assumptions involved may break down in models,  such as the zero-range process, with infinite state space and dynamical phase transitions~\cite{Me05,Me06b}; particular care should be taken in such cases.}   A proof also appears possible at a more rigorous mathematical level by employing older sample path large deviation results of Mogul'skii~\cite{Mogulskii76}.

A natural assumption is that the optimal path minimizing the integral in~\eref{e:mldp} is arranged so that $q(\tau)$ is, in some sense, as close as possible to the temporally local mean current $\bar{j}_{w(q)}$, i.e., the expected current for fixed rates $w(q)$.   Expanding about $\bar{j}_{w(q)}$ and substituting in~\eref{e:mldp} we then have
\begin{equation}
\tilde{I}(j) \approx \lim_{t \to \infty} \min_{q(\tau)} \frac{1}{t^\gamma} \int_{t_0}^t \frac{\left[ q + \tau q' - \bar{j}_{w(q)} \right]^2}{2D_{w(q)}} \, d\tau \label{e:mldp3}
\end{equation}
where $D_{w(q)}$ is the diffusion constant corresponding to rates $w(q)$.  This form clearly reveals the similarity with the spatial additivity result of~\cite{Bodineau04} but it is worth emphasizing that, in the present context, it is only an approximation.  For general final current $j$ it is impossible to find a minimizing path $q(\tau)$ which asymptotically converges to $\bar{j}_{w(q)}$.  Indeed this is already obvious for fluctuations far from the mean in the standard Markovian case.

Whilst~\eref{e:mldp} and~\eref{e:mldp2} may seem to be a powerful general result, their direct application is somewhat limited in practice since, even for those models in which the corresponding Markovian rate function is known, the Euler-Lagrange equations involved in the minimization are typically too complicated to be solved analytically whether or not the form~\eref{e:mldp3} is used.  The known exceptions~\cite{Me09} include various types of history-dependent random walk, including those where the current for left and right jumps is counted separately.  One particularly simple case discussed in~\ref{A:single} is the unidirectional model with rate $v(j)=aj$ which already demonstrates the existence of a large deviation principle with $\gamma$ smaller than unity for a range of ``strong''  memory dependence ($1/2<a<1$).  Physically, this corresponds to a transition to superdiffusive behaviour where the fluctuations of integrated current (equivalently, the position of the random walker) scale faster than linearly with time.  Such a transition was already seen in the elephant random walk and related models~\cite{Schutz04,Hod04,Huillet08b}.

In the next section, we show how these features emerge from a more general approximate analysis which involves an expansion about the fixed points of the dynamics and can easily be applied to complicated many-particle systems.

\section{Fixed point analysis}

\label{s:fixed}

\begin{figure}
\centering
\psfrag{j}[][]{\textcolor{blue}{$q$}}
\psfrag{w2}[Cr][Cr]{\textcolor{red}{$f(q)$}}
\psfrag{w1}[Tr][Br]{\textcolor{red}{$f(q)$}}
\psfrag{stable}[][]{}
\psfrag{unstable}[][]{}
\includegraphics[width=0.7\columnwidth]{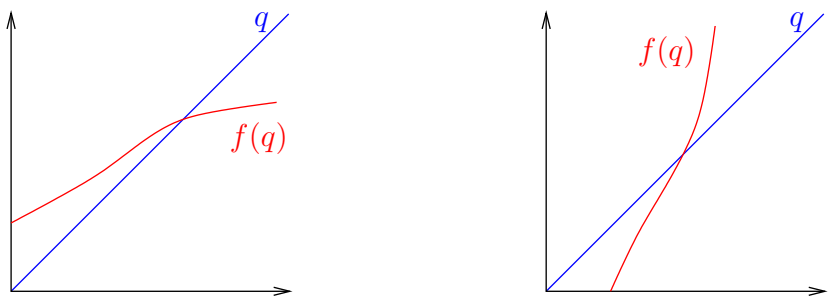}
\caption{Sketch of current fixed points given by the intersection of the function $f(q):=\bar{j}_{w(q)}$ with the diagonal $q$: stable (left) and unstable (right) cases.}
\label{f:stab}
\end{figure}

Lightening the notation by defining $f(q):=\bar{j}_{w(q)}$, it is intuitively clear that a fixed point of the current must obey
\begin{equation}
q=f(q). \label{e:fp}
\end{equation}
In other words, the expected time-averaged current flowing in the next infinitesimal time interval must be the same as that observed in the past.   We denote a fixed point value satisfying~\eref{e:fp} by $j^*$ and now turn to examine its stability which, as illustrated in figure~\ref{f:stab},
is determined by the slope 
\begin{equation}
A^*:=\left. \frac{df}{dq} \right|_{q=j^*}.
\end{equation}
Specifically, if $A^*<1$ (left panel of figure~\ref{f:stab}) then fluctuations above the fixed point yield on average an instantaneous current $f(q)$ which is smaller than the historically-averaged current $q$ and thus there is a reduction back towards the fixed point.  Similarly, a fluctuation below the fixed point has $f(q)>q$ so on average the current increases back towards the fixed point.   Hence, a fixed point with $A^*<1$ is stable and, by the reverse argument, one with $A^*>1$ is unstable (right panel of figure~\ref{f:stab}).\footnote{This argument implicitly assumes that the system decays to stationarity on a timescale which is short compared with the rate of change of the time-averaged current, so that the instantaneous current is well described by $f(j)$.  This is equivalent to the quasistatic assumption of the temporal additivity principle and, at least for finite state space, should always be true for long enough times.}  This heuristic picture, which is essentially the continuous-time version of a ``cobweb'' stability analysis for a discrete mapping, can be made more precise by considering the differential equation for the time-dependence of the expected current.  This latter confirms that decay towards, or growth away from, a fixed point is typically power law in nature which is physically related to the fact that the time-averaged current changes more slowly as time increases.

It is relatively easy to construct models with multiple stable fixed points whose selection is influenced by the early-time behaviour, cf., e.g.,~\cite{Hill80, Mori15, Me15} for the discrete-time case.  This is especially true in the case of non-monotonic current dependence or multiple currents.\footnote{For example, a bidirectional continuous-time random walk in which the hopping rates right and left depend separately on the time-averaged number of jumps right and left as $v_R(j_R,j_L)=aj_R/(j_R+j_L)$ and $v_L(j_R,j_L)=aj_L/(j_R+j_L)$, respectively, has fixed points for $j_R$ and $j_L$ satisfying $-a<j_R-j_L<a$ (with $j_R+j_L=a$).  It can readily be checked, via exact minimization, that the rate function for the net current $j=j_R-j_L$ is zero for the corresponding range of values.} However, in this paper, we specialize to systems in which there is a unique stationary state corresponding to a stable fixed point of the dynamics with some current $j^*$ and slope $A^*$ less than unity.   In this case we can expand $q(\tau)$ about $j^*$ in the numerator and denominator of~\eref{e:mldp3} and keep terms to leading order to obtain 
\begin{equation}
\tilde{I}(j) \approx \lim_{t \to \infty} \min_{q(\tau)} \frac{1}{t^\gamma} \int_{t_0}^t \frac{\left[(1-A^*)(q-j^*)+\tau q' \right]^2}{2D^*} \, d\tau \label{e:gauss}
\end{equation}
where $D^*:=D_{w(j^*)}=(I''_{w(j)}(j)|_{j=j^*})^{-1}$ is assumed non-zero.  
Although we are now guaranteed to get a Gaussian form for $\tilde{I}(j)$ this approach should correctly capture the scaling behaviour of small fluctuations and, in particular, the dependence on $A^*$.  

The minimization in~\eref{e:gauss} is straightforwardly carried out; the corresponding Euler-Lagrange equations are linear and yield an optimal current path of the form
\begin{equation}
q(\tau)=j^*+K_1 \tau^{-A^*} + K_2 \tau^{A^*-1}. \label{e:ELsol}
\end{equation}
Here the integration constants are determined by the boundary conditions ($q(t_0)=j_0$, $q(t)=j$) as
\begin{eqnarray}
K_1&=&\frac{(j_0-j^*)t_0^{1-A^*}-(j-j^*)t^{1-A^*}}{t_0^{1-2A^*}-t^{1-2A^*}} \label{e:bc1} \\
K_2&=&\frac{(j_0-j^*)t_0^{A^*}-(j-j^*)t^{A^*}}{t_0^{2A^*-1}-t^{2A^*-1}} . \label{e:bc2}
\end{eqnarray}
We now substitute~\eref{e:ELsol} into the integrand of~\eref{e:gauss} and carry out the integration to find
\begin{equation}
\int_{t_0}^t {I}_{w(q)}(q+\tau q') \, d\tau = \frac{(1-2A^*)}{2D^*} (K_1)^2 (t^{1-2A^*}-t_0^{1-2A^*}). \label{e:int}
\end{equation}
Finally, inserting the form \eref{e:bc1} for $K_1$ reveals that the right-hand side of~\eref{e:int} scales asymptotically linearly with $t$ (so we need $\gamma=1$ for a non-zero limit) for $A^*< {1}/{2}$, and as $t^{2-2A^*}$ (so $\gamma=2-2A^*$) for $A^* > {1}/{2}$.  To be precise, we end up with a modified large deviation principle of the form
\begin{equation}
\fl
\mathrm{Prob}\left(\frac{\mathcal{J}_t}{t}=j\right)\sim
\cases{
\exp\left[-\frac{(1-2A^*) (j-j^*)^2 }{2D^*}t \right] & for $A^*< \frac{1}{2}$ \\
\exp\left[-\frac{(2A^*-1) (j-j^*)^2}{2D^*}t_0^{2A^*-1}t^{2-2A^*} \right] & for $A^* > \frac{1}{2}$. \\
} \label{e:scale}
\end{equation}
Physically, for $A^* < {1}/{2}$, there is diffusive behaviour with a modified diffusion coefficient $D^*/(1-2A^*)$.  We see clearly here that $A^*$ quantifies the effective strength of the feedback -- for $A^*$ negative, fluctuations are suppressed whilst, for $A^*$ positive, they are enhanced.  For $A^* > {1}/{2}$, there is superdiffusive behaviour which retains an ageing-type dependence on the initial time $t_0$.  At $A^*=1/2$ one expects logarithmic corrections whose analysis is beyond the scope of the current paper.

As mentioned earlier, this transition is consistent with that already observed in the single-particle example of~\ref{A:single} and other random walk models~\cite{Schutz04,Hod04,Huillet08b,Me15}.  In the next section we will illustrate the power of the general approach by appeal to a specific many-particle system.

\section{Exclusion process with current-dependent memory}
\label{s:ASEP}

\subsection{Model}
The totally asymmetric simple exclusion process (TASEP) was first introduced in 1968 to describe protein synthesis~\cite{MacDonald68} and, since then, has enjoyed widespread success both as a base model for various transport processes~\cite{Chowdhury00,Chowdhury05b} and as a vehicle for advancing theoretical understanding of non-equilibrium systems see, e.g.,~\cite{Derrida98c,Golinelli06b,Chou11} and references therein.   We here start from the standard continuous-time version of this model on a one-dimensional lattice with open boundaries and modify it to include a current-dependent input rate.  Obviously, many other forms of current dependence could be envisaged but this is a natural choice as a form of feedback -- the reader is invited to imagine controlling the arrival of cars onto a stretch of road.

\begin{figure}
\centering
\psfrag{a}{\textcolor{red}{$\alpha(j)$}}
\psfrag{b}{$\beta$}
\psfrag{1}[][]{1}
\psfrag{2}[][]{2}
\psfrag{3}[][]{3}
\psfrag{l-1}[][]{$l\!-\!1$}
\psfrag{l}[][]{$l$}
\psfrag{l+1}[][]{$l\!+\!1$}
\psfrag{L}[][]{$L$}
\psfrag{p}[][]{$p$}
\psfrag{ql}[][]{$q$}
\includegraphics*[width=0.8\textwidth]{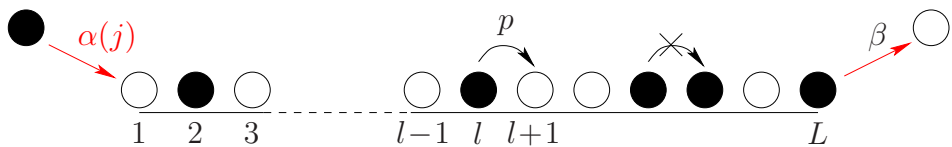}
\caption{Schematic of one-dimensional TASEP with input rate depending on time-averaged current $j$ over the whole past history.}
\label{f:ASEPj}
\end{figure}

To be more concrete, our model is defined in the following manner (see also figure~\ref{f:ASEPj}).  Each of the $L$ lattice sites has only two possible configurations: occupied (particle) or vacant (hole).  A particle on site $l$ hops after an exponentially distributed waiting time with mean $1/p$ to site $l+1$ if, and only if, that site is vacant.  Without loss of generality, we set the rate $p=1$ in the following.  Particles are removed at the right-hand boundary (site $L$) with rate $\beta$ and injected subject to the exclusion rule at the left-hand boundary (site $1$) with a rate $\alpha(j)$ which, crucially, is a function of the time-averaged input current over the whole previous history.  In fact, it is obvious from the continuity equation that, for a finite chain in the long-time limit, the time-averaged current must be the same between any pair of nearest-neighbour sites.

For illustrative purposes (and in analogy with the single-particle analysis of~\ref{A:single}) we mainly consider a linear current dependence of the form
\begin{equation}
\alpha(j)=\alpha_0 + aj, \label{e:lin} 
\end{equation}
where $0 \leq \alpha_0 \leq 1$ and $a>0$, before later touching on some other choices.  An input rate of apparently similar form to~\eref{e:lin} was independently proposed by Sharma and Chowdhury~\cite{Sharma11b} to model the recycling of ribosomes in protein synthesis and implemented for the more general case of the $l$-TASEP with extended objects.  We remark here that in~\cite{Sharma11b} one has the restriction $a \leq 1$ (as befits the biological context) and also, significantly, $j$ is the instantaneous mean (output) current rather than the average over the whole previous history.  The relevance of these distinctions should become apparent in the discussion of phase diagrams and fluctuations below.

\subsection{Mean current}

It is well known (see, e.g.,~\cite{Derrida98c}) that, in the thermodynamic limit, the standard Markovian TASEP has the following three regimes.
\begin{itemize}
\item For $\alpha<1/2$, $\beta>\alpha$ there is a low-density (LD) phase in which the mean current is controlled by the input rate and given by $\alpha(1-\alpha)$.
\item For $\alpha>\beta$, $\beta<1/2$ there is a corresponding high density (HD) phase in which the mean current is controlled by the output rate and given by $\beta(1-\beta)$.
\item For $\alpha>1/2$, $\beta>1/2$ the system is in the maximal current (MC) phase where the mean current is limited by the bulk hopping rate and given simply by 1/4.
\end{itemize}
We now seek to determine the effect of the current-dependent memory on the phase boundaries and the mean current in each phase.  

Following the approach of the previous section, we argue that the mean current in the long-time limit is given by the fixed points in the three different regimes:
\begin{equation}
j^* = 
\cases{
\alpha(j^*)(1-\alpha(j^*)) & for $\alpha(j^*) < \frac{1}{2}, \beta > \alpha(j^*)$ [LD]\\
\beta(1-\beta) & for $\alpha(j^*) > \beta, \beta < \frac{1}{2}$ [HD] \\
\frac{1}{4} & for $\alpha(j^*) > \frac{1}{2}, \beta > \frac{1}{2}$ [MC]. \\
} \label{e:TASEPfp}
\end{equation}
Unsurprisingly, since the current dependence is in the input rate, the fixed point is unchanged in HD and MC phases.  In the LD phase, however, some simple algebra yields
\begin{equation}
j^*=\frac{-(2\alpha_0a+1-a)+\sqrt{4\alpha_0 a+(1-a)^2}}{2 a^2}  \label{e:fpld}
\end{equation}
with the other solution to the quadratic corresponding to an unphysical negative current.  

We can readily show that at the value of $j^*$ given by~\eref{e:fpld}    
\begin{equation}
A^*=\left.\frac{d}{dj}\left[\alpha(j)(1-\alpha(j))\right]\right|_{j=j^*}=1-\sqrt{4\alpha_0 a+(1-a)^2}. \label{e:Astar}
\end{equation}
For $0<\alpha_0<{1}/{2}-{a}/{4}$ we have $0<A^*<1$ so this is a stable fixed point with ``positive'' feedback.  Here the upper bound
\begin{equation}
\alpha_0=\frac{1}{2}-\frac{a}{4}
\end{equation}
corresponds to the LD-MC phase transition (determined by $\alpha(j)=1/2$).  Furthermore the LD-HD transition line ($\beta=\alpha(j^*)$) becomes curved rather than straight and is given by
\begin{equation}
\beta=\frac{-(1-a)+\sqrt{4\alpha_0 a+(1-a)^2}}{2a}.
\end{equation}

\begin{figure}
\centering
\psfrag{a}[Tc][Tc]{$\alpha_0$}
\psfrag{b}[Cr][Cr]{$\beta$}
\psfrag{0b}[Tc][Tc]{\scriptsize{0}}
\psfrag{1b}[Tc][Tc]{\scriptsize{1}}
\psfrag{0}[Cr][Cr]{\scriptsize{0}}
\psfrag{1}[Cr][Cr]{\scriptsize{1}}
\psfrag{LD}[Cc][Cc][1][90]{\scriptsize{LD}}
\psfrag{HD}[Cc][Cc]{\scriptsize{HD}}
\psfrag{MC}[Cc][Cc]{\scriptsize{MC}}
\subfigure[$a=0.8$]{\includegraphics[width=0.32\textwidth]{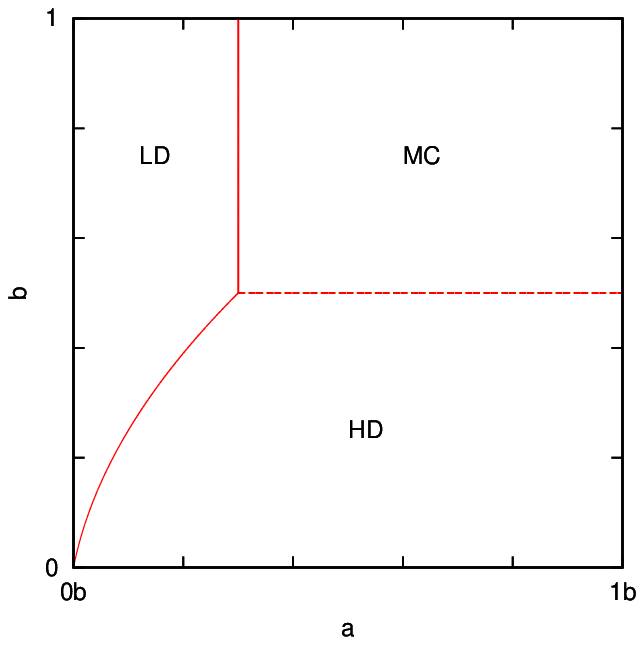}} \hfill
\subfigure[$a=1.6$]{\includegraphics[width=0.32\textwidth]{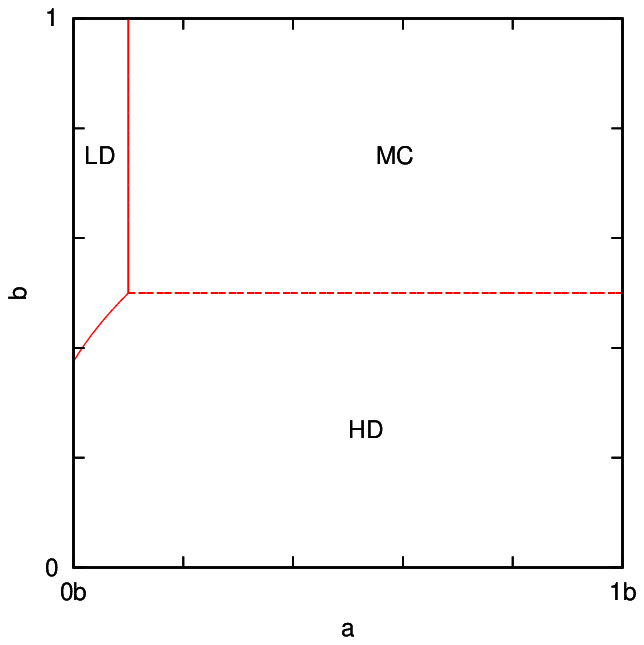}} \hfill
\subfigure[$a=2.4$]{\includegraphics[width=0.32\textwidth]{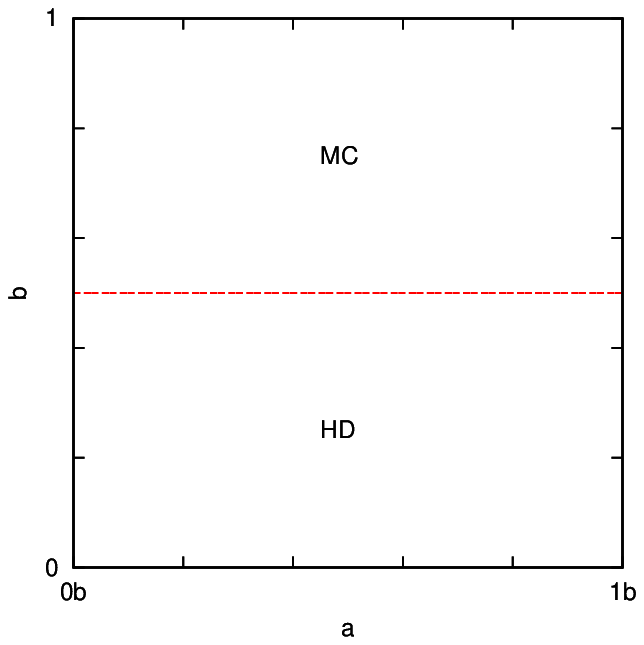}} \hfill
\caption{Phase diagrams for current-dependent TASEP with $\alpha(j)=\alpha_0+aj$ and different values of $a$.  Note that the picture in (c) is unchanged for all $a \geq 2$.}
\label{f:pd}
\end{figure}

As might be intuitively expected, the general effect of the positive feedback resulting from the $aj$ term is to increase the size of the maximal current phase.  However, as exemplified by the representative cases in figure~\ref{f:pd}, 
we predict the following three qualitatively different forms of phase diagram depending on the value of $a$. 
\begin{itemize}
\item For $0 \leq a \leq 1$, the phase diagram reproduces that given in~\cite{Sharma11b} -- the distinction between dependence on historically-averaged and instantaneous current is irrelevant for calculation of the fixed point although not for the fluctuations (next subsection).   The LD-HD transition line always passes through the origin and the phase diagram reduces to the Markovian case when $a=0$.
\item For $1<a<2$ there is a qualitative difference in that the LD-HD phase transition line intersects the $\beta$ axis at $\beta>0$.    The feedback is strong enough to ensure a non-zero mean-current in the LD phase even for $\alpha_0 \to 0$; at $\alpha_0=0$ there is an unstable fixed point at zero and a stable fixed point at $j^*=(a-1)/a^2$.
\item For $a \geq 2$ there is no LD phase.  In other words, $\alpha_0$ never controls the current -- for $\beta<1/2$ it is determined by the output rate and for $\beta>1/2$ by the bulk hopping rate.
\end{itemize}

\begin{figure}
\centering
\psfrag{a}[Cc][Cc]{$\alpha_0$}
\psfrag{b}[Cc][Cc]{$\beta$}
\psfrag{j}[Cc][Cc]{$\mathcal{J}/t$}
\psfrag{0.0}[Cr][Cr]{\scriptsize{0.0}}
\psfrag{0.1}[Cr][Cr]{\scriptsize{0.1}}
\psfrag{0.2}[Cr][Cr]{\scriptsize{0.2}}
\psfrag{0.3}[Cr][Cr]{\scriptsize{0.3}}
\psfrag{0.4}[Cr][Cr]{\scriptsize{0.4}}
\psfrag{0.5}[Cr][Cr]{\scriptsize{0.5}}
\psfrag{0.6}[Cr][Cr]{\scriptsize{0.6}}
\psfrag{0.7}[Cr][Cr]{\scriptsize{0.7}}
\psfrag{0.8}[Cr][Cr]{\scriptsize{0.8}}
\psfrag{0.9}[Cr][Cr]{\scriptsize{0.9}}
\psfrag{1.0}[Cr][Cr]{\scriptsize{1.0}}
\psfrag{0.00}[Cl][Cl]{\scriptsize{0.00}}
\psfrag{0.05}[Cl][Cl]{\scriptsize{0.05}}
\psfrag{0.10}[Cl][Cl]{\scriptsize{0.10}}
\psfrag{0.15}[Cl][Cl]{\scriptsize{0.15}}
\psfrag{0.20}[Cl][Cl]{\scriptsize{0.20}}
\psfrag{0.25}[Cl][Cl]{\scriptsize{0.25}}
\includegraphics[width=0.8\textwidth]{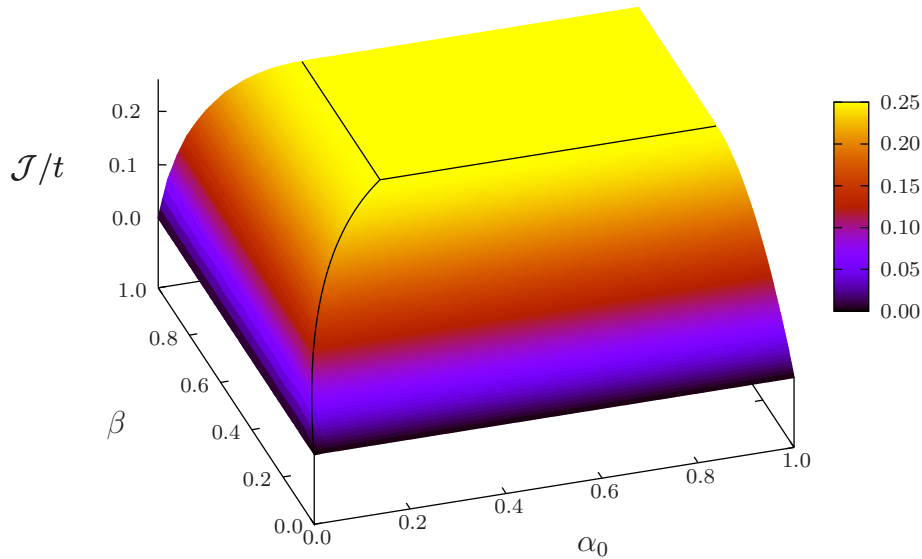}
\caption{Monte Carlo simulation data for final value of time-averaged current $\mathcal{J}/t$ as a function of rates $\alpha_0$ and $\beta$ for a single trajectory of length $t=10^6$ in a system of size $L=1000$ with $\alpha(j)=\alpha_0+0.8j$.  Initial condition used was $t_0=1$, $j_0=0$, and each site independently occupied by a particle with probability corresponding to the bulk density of a Markovian TASEP with the same input and output rates; for times $\tau>t_0$ a discrete-time random sequential update rule was used with 20 steps per unit time up to $\tau=1000$
(allowing for the fact that the time-averaged current $q(\tau)$ changes relatively fast at the beginning of the trajectory) and 2 steps per unit time thereafter.  Data sampled at boundary rate increments of 0.02 and interpolated with gnuplot.  Solid black lines show theoretically predicted phase boundaries.}
\label{f:pdsim}
\end{figure}

\begin{figure}
\centering
\psfrag{a}[Tc][Tc]{$\alpha_0$}
\psfrag{j}[Cc][Cc]{${\langle \mathcal{J}\rangle}/{t}$}
\psfrag{0.0}[Tc][Tc]{\scriptsize{0.0}}
\psfrag{0.2}[Tc][Tc]{\scriptsize{0.2}}
\psfrag{0.4}[Tc][Tc]{\scriptsize{0.4}}
\psfrag{0.6}[Tc][Tc]{\scriptsize{0.6}}
\psfrag{0.8}[Tc][Tc]{\scriptsize{0.8}}
\psfrag{1.0}[Tc][Tc]{\scriptsize{1.0}}
\psfrag{0.00}[Cr][Cr]{\scriptsize{0.00}}
\psfrag{0.05}[Cr][Cr]{\scriptsize{0.05}}
\psfrag{0.10}[Cr][Cr]{\scriptsize{0.10}}
\psfrag{0.15}[Cr][Cr]{\scriptsize{0.15}}
\psfrag{0.20}[Cr][Cr]{\scriptsize{0.20}}
\psfrag{0.25}[Cr][Cr]{\scriptsize{0.25}}
\psfrag{0.30}[Cr][Cr]{\scriptsize{0.30}}
\includegraphics[width=0.8\textwidth]{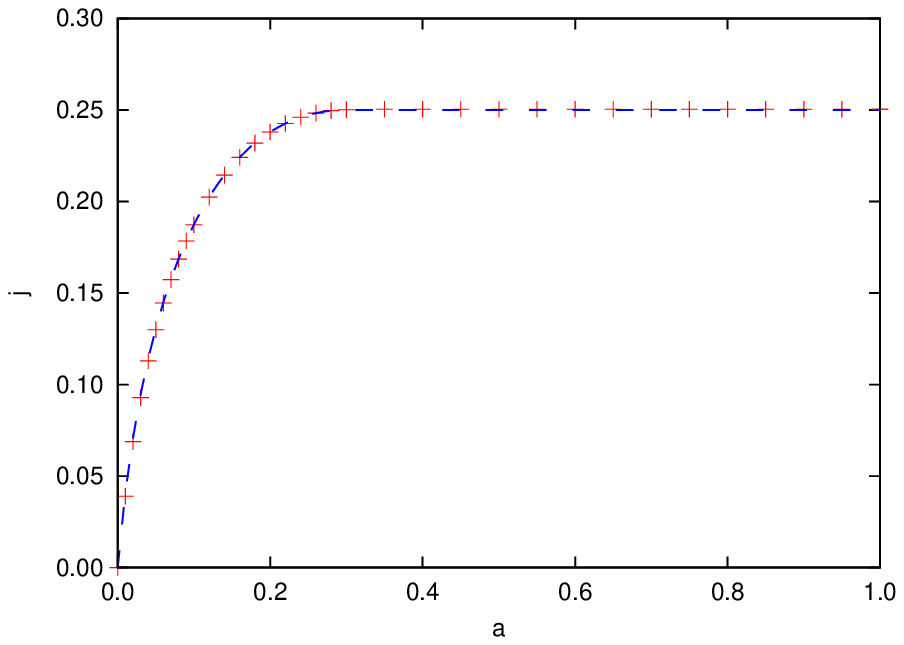}
\caption{Monte Carlo simulation data for mean $\langle \mathcal{J} \rangle /t$ as a function of $\alpha_0$ with $\alpha(j)=\alpha_0+0.8j$ and $\beta=0.6$.  Data from 1000 trajectories of length $t=10^6$ in a system of size $L=1000$.  Other simulation details same as those used for figure~\ref{f:pdsim}.  Blue dashed line is theoretical prediction of~\eref{e:fpld} for $j^*$.}
\label{f:cross}
\end{figure}

The fixed points in the different regimes of these phase diagrams are confirmed by Monte Carlo simulations. 
For example, in figure~\ref{f:pdsim} we show a three-dimensional plot of the final current for a single long trajectory  (as a function of boundary rates $\alpha_0$ and $\beta$) in the model with $a=0.8$ and $L=1000$; the accord with the theoretically predicted phase boundaries is self-evident. 
As a further check, we then plot in figure~\ref{f:cross} the mean time-averaged current from 1000 different histories for the cross-section of the phase diagram with $\beta=0.6$.  
The quantitative agreement of the mean current with the predicted fixed point $j^*$~\eref{e:fpld} is very good and similar confirmation is found for other rate parameters.  However, due to the size of the system, one does need to simulate for relatively long times until the rate of change of the current is slow compared to the decay to stationarity and the quasistatic assumption is reasonable.\footnote{Formally the method requires that the long-time $t \to \infty$ limit is taken \emph{before} the thermodynamic $L \to \infty$ limit; this is also important for the study of fluctuations in the next section.}  For smaller systems, the decay to stationarity is obviously faster but there are finite-size corrections for the mean currents~\cite{Derrida93b} which would require $L$-dependent expressions on the right-hand side of \eref{e:TASEPfp} and, in general, numerical solution for the LD fixed point.   For different initial conditions the current may be different for short times but should eventually approach the same stable fixed point except for in the special case where the system is started exactly at an unstable fixed point.  This latter is relevant for simulations at $\alpha_0=0$ in the $a>1$ case where an initial condition of $j_0=0$ is observed to lead to a zero current for all times whereas $j_0>0$ gives convergence to the stable fixed point $j^*=(a-1)/a^2$. 

In concluding this subsection we note that modified phase diagrams have been calculated for many other variants of the TASEP including those with stochastic gating (which can be thought of as the introduction of additional ``hidden'' variables in the standard Markovian model)~\cite{Wood09} and density feedback control~\cite{Woelki13}.  However, we stress here that our approach enables us not only to predict the mean current but also to gain information about the fluctuations, as we shall see in the next subsection.

\subsection{Fluctuations}

According to the analysis of the Markovian TASEP in~\cite{Derrida95}, the diffusion constant in the MC phase scales asymptotically as $L^{-1/2}$ so, in the thermodynamic limit, $D^* \to 0$ and~\eref{e:gauss} is not applicable.  However, in both HD and LD phases the diffusion constant approaches a finite limit -- here we aim to understand the effect of the memory on the fluctuations in the latter case.  

\begin{figure}
\centering
\psfrag{t}[Tc][Tc]{$t$}
\psfrag{s}[Cc][Cc]{$\mathrm{Var}(\mathcal{J}/t)$}
\psfrag{ 1000}[Tc][Tc]{\scriptsize{$10^3$}}
\psfrag{ 10000}[Tc][Tc]{\scriptsize{$10^4$}}
\psfrag{ 100000}[Tc][Tc]{\scriptsize{$10^5$}}
\psfrag{ 1e+06}[Tc][Tc]{\scriptsize{$10^6$}}
\psfrag{ 0.1}[Tc][Tc]{\scriptsize{$10^{-1}$}}
\psfrag{ 0.01}[Tc][Tc]{\scriptsize{$10^{-2}$}}
\psfrag{ 0.001}[Tc][Tc]{\scriptsize{$10^{-3}$}}
\psfrag{ 0.0001}[Cr][Cr]{\scriptsize{$10^{-4}$}}
\psfrag{ 1e-05}[Cr][Cr]{\scriptsize{$10^{-5}$}}
\psfrag{ 1e-06}[Cr][Cr]{\scriptsize{$10^{-6}$}}
\psfrag{ 1e-07}[Cr][Cr]{\scriptsize{$10^{-7}$}}
\psfrag{ 1e-08}[Cr][Cr]{\scriptsize{$10^{-8}$}}
\includegraphics[width=0.8\textwidth]{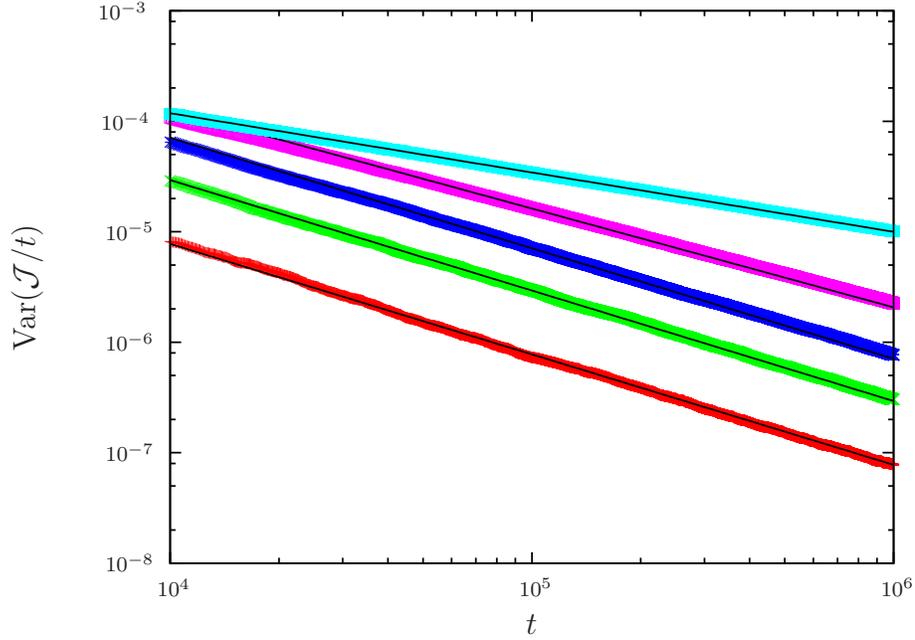}
\caption{Variance of $\mathcal{J}/t$ as a function of time for selected values of $\alpha_0$ with other parameters as in figure~\ref{f:cross} ($a=0.8$, $\beta=0.6$).  Points show simulation data for (top to bottom): $\alpha_0 = 0.01, 0.05, 0.08, 0.12, 2$. Black solid
lines are fits corresponding to power laws with negative exponent $\min(1, 2 - 2A^*)$; logarithmic corrections are expected at the critical point $\alpha_c= 0.065625$.}
\label{f:scale}
\end{figure}

\begin{figure}
\centering
\psfrag{a}[Tc][Tc]{$\alpha_0$}
\psfrag{j}[Cc][Cc]{$\mathrm{Var}(\mathcal{J})/t$, $\mathrm{Var}(\mathcal{J})/t^{2A^*}$}
\psfrag{0.0}[Tc][Tc]{\scriptsize{0.0}}
\psfrag{0.2}[Tc][Tc]{\scriptsize{0.2}}
\psfrag{0.4}[Tc][Tc]{\scriptsize{0.4}}
\psfrag{0.6}[Tc][Tc]{\scriptsize{0.6}}
\psfrag{0.8}[Tc][Tc]{\scriptsize{0.8}}
\psfrag{1.0}[Tc][Tc]{\scriptsize{1.0}}
\psfrag{0.00}[Cr][Cr]{\scriptsize{0.0}}
\psfrag{0.20}[Cr][Cr]{\scriptsize{0.2}}
\psfrag{0.40}[Cr][Cr]{\scriptsize{0.4}}
\psfrag{0.60}[Cr][Cr]{\scriptsize{0.6}}
\psfrag{0.80}[Cr][Cr]{\scriptsize{0.8}}
\psfrag{1.00}[Cr][Cr]{\scriptsize{1.0}}
\includegraphics[width=0.8\textwidth]{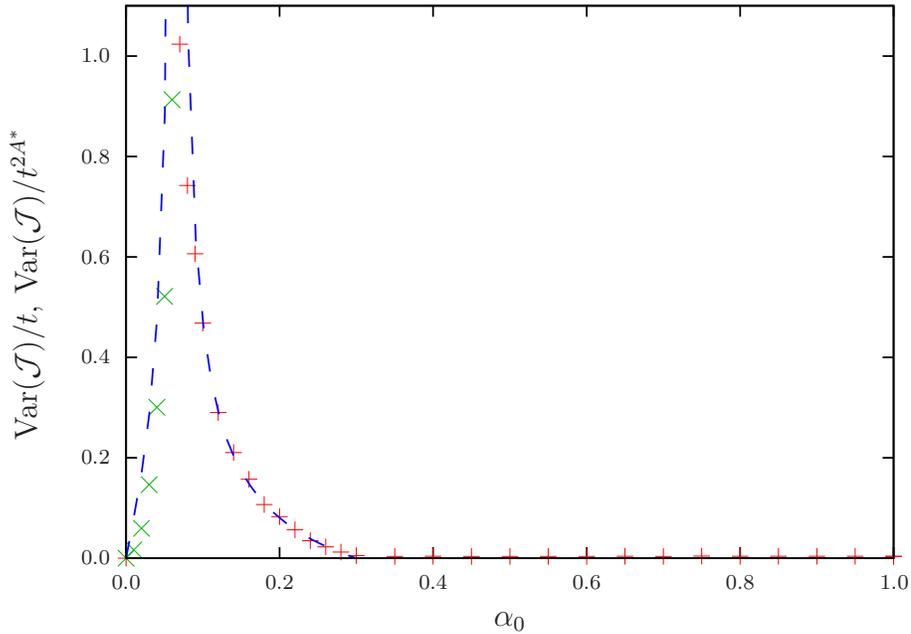}
\caption{Monte Carlo simulation data for variance of $\mathcal{J}$ as a function of $\alpha_0$  for parameters of figure~\ref{f:cross} ($a=0.8$, $\beta=0.6$).  Red $+$ symbols show $(\langle\mathcal{J}^2\rangle-\langle\mathcal{J}\rangle^2)/t$, expected to have finite long-time limit in diffusive regime ($\alpha > \alpha_c = 0.065625$); green $\times$ symbols show $(\langle\mathcal{J}^2\rangle-\langle\mathcal{J}\rangle^2)/t^{2A^*}$, predicted to be finite in superdiffusive regime ($\alpha < \alpha_c$).  Blue dashed line is theoretical prediction for $D^*/|1-2A^*|$ from~\eref{e:Astar} and~\eref{e:Dstar}.}
\label{f:super}
\end{figure}

Starting from the Markovian result in~\cite{Derrida95} we have, for the LD phase,
\begin{equation}
D^*=\alpha(j^*)(1-\alpha(j^*))(1-2\alpha(j^*)) \label{e:Dstar}
\end{equation}
where, for our model with $\alpha(j)=\alpha_0+aj$, the fixed point $j^*$ is given by~\eref{e:fpld}.  Now, as argued in section~\ref{s:fixed}, we expect long-time diffusive behaviour with modified diffusion coefficient  $D^*/(1-2A^*)$ when $A^*$ of~\eref{e:Astar} is less than 1/2.  Significantly, however there should be long-time superdiffusive behaviour for $1/2 < A^* < 1$ which is true for
\begin{equation}
\alpha_0 < \frac{{1}/{4}-(1-a)^2}{4a} =: \alpha_c.
\end{equation}
Note that $\alpha_c$ is positive only for $1/2<a<3/2$; in that range, we predict a subregime in the LD phase for which the fluctuations are superdiffusive and the variance of the time averaged current $\mathcal{J}/t$ scales in the long-time limit as $t^{2A^*-2}$.  This asymptotic scaling is fairly convincingly supported by a log-log plot of variance against time for selected values of $\alpha_0$ in the $a=0.8$ case (figure~\ref{f:scale}).
Additionally, figure~\ref{f:super} shows a naive check on the predicted coefficient $D^*/|1-2A^*|$ across a constant-$\beta$ cross-section of the phase diagram with the divergence at $\alpha_c$ clearly to be seen.
The slight theoretical overestimation of the $t=10^6$ data for small $\alpha_0$ (corresponding to small mean current)  may be related to the fact that, for totally asymmetric systems such as this, the current distribution must be cut off at $j=0$ and the Gaussian approximation is thus expected to be less good for small $j^*$ (and inapplicable for $j^*=0$).

For the same model with $a=1.6$, preliminary simulations (not shown) support the assertion that there is no superdiffusion and, in fact, suggest that the width of the time-averaged current distribution decays somewhat faster than the diffusive prediction of~\eref{e:scale}, at least for intermediate times.  Again this may be related to the $j=0$ cut-off but further investigation for longer times would certainly be desirable.   More generally, the existence of a superdiffusive subregime in the phase diagram clearly depends on the precise form of the current dependence.  For example, in another tractable case $\alpha(j)=\alpha_0 + a\sqrt{j}$ we also see the MC phase extended at the expense of the LD phase but predict that fluctuations throughout the LD phase remain diffusive for all values of $a$.  In this case too, the mean current and absence of superdiffusion are confirmed by simulation but more work is still needed to definitively determine the applicability of~\eref{e:scale}.

\subsection{Exact numerical minimization}

Going beyond the mean and diffusion coefficient, the current large deviations in the Markovian TASEP have remarkably been fully characterized recently for all hopping rates and systems sizes~\cite{DeGier11b,Lazarescu11b,Gorissen12b}.  We can now use these results to evaluate~\eref{e:mldp} directly and thus to check the consistency of the Gaussian approximation applied in the previous subsections.  

In the LD phase the scaled cumulant generating function (the Legendre transform of the rate function) approaches an $L$-independent limit as the system size increases.  Specifically, we have
\begin{equation}
\fl \lim_{t \to \infty} \frac{1}{t} \log \langle e^{-\lambda \mathcal{J}} \rangle = \alpha(1-\alpha)\left( \frac{1-e^{-\lambda}}{1-\alpha+\alpha e^{-\lambda}} \right) \quad \textrm{for}~~-\log\left(\frac{1-\alpha}{\alpha}\right) < \lambda < \infty 
\label{e:TASEPscgf}
\end{equation}
which straightforwardly corresponds to
\begin{equation}
\fl I_\alpha(j)=\frac{(2\alpha-1)+\sqrt{1-4j}}{2} + j \log \left[ \frac{(1-\alpha)(1-2j-\sqrt{1-4j})}{2 \alpha j} \right]\quad \textrm{for}~~0 < j < 1/4.
\label{e:TASEPrf}
\end{equation}
The form~\eref{e:TASEPscgf} was obtained by Bethe ansatz in~\cite{DeGier11b} and via a general parametric representation in~\cite{Lazarescu11b}.  It can also be derived within the framework of macroscopic fluctuation theory~\cite{Bodineau06}.  One can readily check that the rate function~\eref{e:TASEPrf} has a zero at the mean current $\bar{j}_\alpha=\alpha(1-\alpha)$ with (inverse) second derivative at that point corresponding to the diffusion coefficient $\alpha(1-\alpha)(1-2\alpha)$.   At $j = 1/4$ there is a dynamical phase transition to a regime in which $I_\alpha(j)$ retains a dependence on $L$.  Nevertheless, at least away from this transition we claim that the rate function of the non-Markovian current-dependent process should be given by minimizing the integral in~\eref{e:mldp} with an integrand $I_{\alpha(q)}(q+\tau q')$ which is simply obtained from~\eref{e:TASEPrf} via the replacement of $\alpha$ with $\alpha(q)$.  In practice, this integral is much too complicated to approach analytically so we resort to exact numerical calculations using Mathematica.   Some computational difficulties are encountered here, apparently related to stiffness of the differential equations (as well as perhaps the finite range of applicability for $I_\alpha(j)$ and the impossibility of negative currents).  However, notwithstanding this, the approach enables us to push the bounds of investigation beyond the Gaussian regime discussed above.

\begin{figure}
\centering
\psfrag{0.00b}[Tc][Tc]{\scriptsize{0.00}}
\psfrag{0.05b}[Tc][Tc]{\scriptsize{0.05}}
\psfrag{0.10b}[Tc][Tc]{\scriptsize{0.10}}
\psfrag{0.15b}[Tc][Tc]{\scriptsize{0.15}}
\psfrag{0.20b}[Tc][Tc]{\scriptsize{0.20}}
\psfrag{0.25b}[Tc][Tc]{\scriptsize{0.25}}
\psfrag{0.00}[Cr][Cr]{\scriptsize{0.00}}
\psfrag{0.02}[Cr][Cr]{\scriptsize{0.02}}
\psfrag{0.04}[Cr][Cr]{\scriptsize{0.04}}
\psfrag{0.06}[Cr][Cr]{\scriptsize{0.06}}
\psfrag{0.08}[Cr][Cr]{\scriptsize{0.08}}
\psfrag{0.10}[Cr][Cr]{\scriptsize{0.10}}
\psfrag{0.12}[Cr][Cr]{\scriptsize{0.12}}
\psfrag{0.14}[Cr][Cr]{\scriptsize{0.14}}
\psfrag{0.16}[Cr][Cr]{\scriptsize{0.16}}
\psfrag{0.18}[Cr][Cr]{\scriptsize{0.18}}
\psfrag{0.10i}[Tc][Tc]{\scriptsize{0.10}}
\psfrag{0.11i}[Tc][Tc]{\scriptsize{0.11}}
\psfrag{0.12i}[Tc][Tc]{\scriptsize{0.12}}
\psfrag{0.13i}[Tc][Tc]{\scriptsize{0.13}}
\psfrag{0.14i}[Tc][Tc]{\scriptsize{0.14}}
\psfrag{0.15i}[Tc][Tc]{\scriptsize{0.15}}
\psfrag{0.16i}[Tc][Tc]{\scriptsize{0.16}}
\psfrag{0.17i}[Tc][Tc]{\scriptsize{0.17}}
\psfrag{0.18i}[Tc][Tc]{\scriptsize{0.18}}
\psfrag{0.19i}[Tc][Tc]{\scriptsize{0.19}}
\psfrag{0.20i}[Tc][Tc]{\scriptsize{0.20}}
\psfrag{0.000}[Cr][Cr]{\scriptsize{0.000}}
\psfrag{0.002}[Cr][Cr]{\scriptsize{0.002}}
\psfrag{0.004}[Cr][Cr]{\scriptsize{0.004}}
\psfrag{0.006}[Cr][Cr]{\scriptsize{0.006}}
\psfrag{0.001}[Cr][Cr]{\scriptsize{}}
\psfrag{0.003}[Cr][Cr]{\scriptsize{}}
\psfrag{0.005}[Cr][Cr]{\scriptsize{}}
\psfrag{0.007}[Cr][Cr]{\scriptsize{}}
\psfrag{0.008}[Cr][Cr]{\scriptsize{0.008}}
\psfrag{0.009}[Cr][Cr]{\scriptsize{}}
\psfrag{j}[Tc][Tc]{$j$}
\psfrag{I}[Tc][Tc]{$\tilde{I}(j,1000)$}
\psfrag{ji}[Tc][Tc]{}
\psfrag{Ii}[Tc][Tc]{}
\def\stackalignment{r}
\topinset{\includegraphics[width=0.45\textwidth]{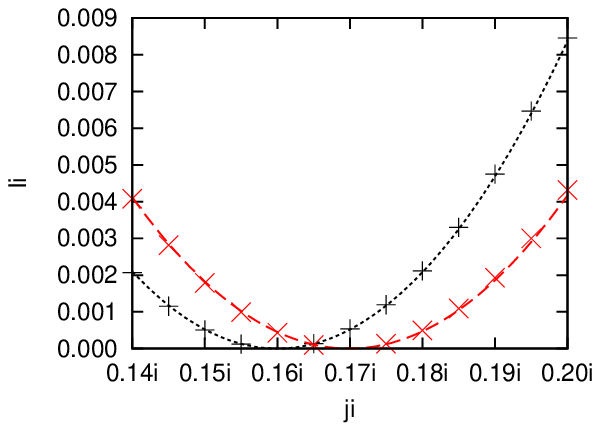}}{\includegraphics[width=0.8\textwidth]{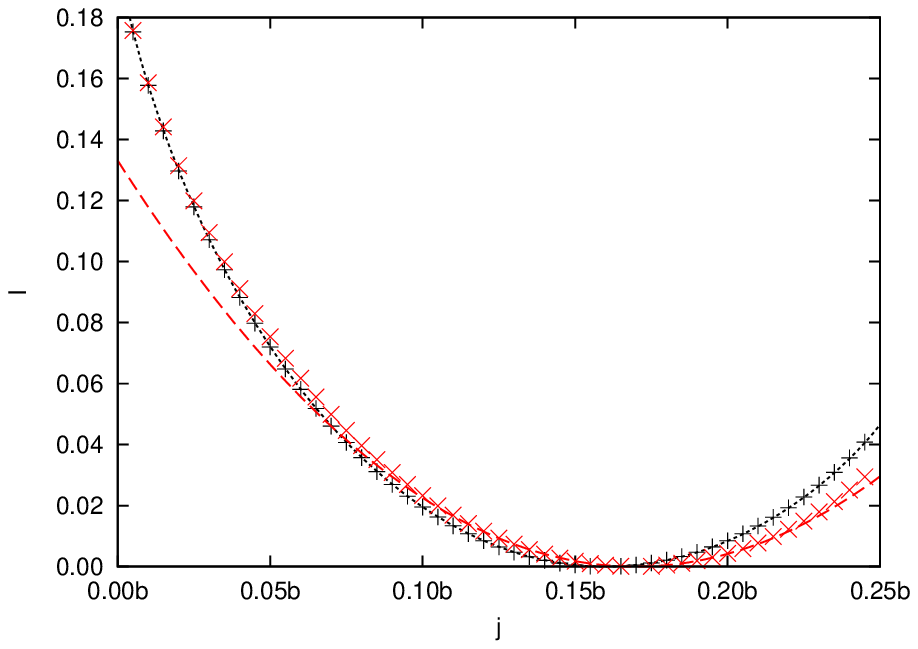}}{0.05\textwidth}{0.10\textwidth}
\caption{Mathematica results for LD-phase $\tilde{I}(j,1000)$ in case $\alpha(j)=\alpha_0 + a j$ with $\alpha_0=0.2$,  $a=0$ (black $+$ symbols) and $a=0.1$ (red $\times$ symbols); initial condition $t_0=1$, $j_0=0$.  Black dotted line is exact expression~\eref{e:TASEPrf} for Markovian rate function $I_{\alpha_0}(j)$; red dashed line is Gaussian expansion of non-Markovian $\tilde{I}(j)$ for $a=0.1$.  Inset shows close-up around mean current.}
\label{f:numerics}
\end{figure}

Returning to our favourite example with $\alpha(j)=\alpha_0 + a j$, we focus now on small values of $a$ because they lead to more stable numerics and yet clearly illustrate the effect of even weak memory dependence on the current large deviations.  Figure~\ref{f:numerics} shows the finite-time quantity
\begin{equation}
\tilde{I}(j,t) = \min_{q(\tau)} \frac{1}{t} \int_{t_0}^t {I}_{\alpha(q)}(q+\tau q') \, d\tau
\end{equation}
evaluated at $t=1000$ for fixed $\alpha_0$ and both zero and non-zero values of $a$. 
In the $a=0$ case we find excellent agreement with the Markovian rate function~\eref{e:TASEPrf} and we anticipate that our numerical method converges fast towards the long-time limit also in the non-Markovian case.   For $a>0$ the mean current (zero of the rate function) is shifted to a higher value and the width of the distribution increased.  The approximation from the fixed point analysis of the preceding subsections matches very well the behaviour for small fluctuations but, as expected, is inaccurate for larger fluctuations.  In particular, by construction, the Gaussian fails to capture the asymmetry of the rate function about the mean --  we need the full minimization to see that the probability of large fluctuations below the mean is hardly affected by the memory whereas large fluctuations above the mean become much more likely than in the Markovian case (presumably because, in this model, the feedback can increase but never decrease the hopping rate). 

\begin{figure}
\centering
\psfrag{0.00b}[Tc][Tc]{\scriptsize{0.00}}
\psfrag{0.05b}[Tc][Tc]{\scriptsize{0.05}}
\psfrag{0.10b}[Tc][Tc]{\scriptsize{0.10}}
\psfrag{0.15b}[Tc][Tc]{\scriptsize{0.15}}
\psfrag{0.20b}[Tc][Tc]{\scriptsize{0.20}}
\psfrag{0.25b}[Tc][Tc]{\scriptsize{0.25}}
\psfrag{0.00}[Cr][Cr]{\scriptsize{0.00}}
\psfrag{0.02}[Cr][Cr]{\scriptsize{0.02}}
\psfrag{0.04}[Cr][Cr]{\scriptsize{0.04}}
\psfrag{0.06}[Cr][Cr]{\scriptsize{0.06}}
\psfrag{0.08}[Cr][Cr]{\scriptsize{0.08}}
\psfrag{0.10}[Cr][Cr]{\scriptsize{0.10}}
\psfrag{0.12}[Cr][Cr]{\scriptsize{0.12}}
\psfrag{0.14}[Cr][Cr]{\scriptsize{0.14}}
\psfrag{0.10i}[Tc][Tc]{\scriptsize{0.10}}
\psfrag{0.11i}[Tc][Tc]{\scriptsize{0.11}}
\psfrag{0.12i}[Tc][Tc]{\scriptsize{0.12}}
\psfrag{0.13i}[Tc][Tc]{\scriptsize{0.13}}
\psfrag{0.14i}[Tc][Tc]{\scriptsize{0.14}}
\psfrag{0.15i}[Tc][Tc]{\scriptsize{0.15}}
\psfrag{0.16i}[Tc][Tc]{\scriptsize{0.16}}
\psfrag{0.000}[Cr][Cr]{\scriptsize{0.000}}
\psfrag{0.002}[Cr][Cr]{\scriptsize{0.002}}
\psfrag{0.004}[Cr][Cr]{\scriptsize{0.004}}
\psfrag{0.006}[Cr][Cr]{\scriptsize{0.006}}
\psfrag{0.001}[Cr][Cr]{\scriptsize{}}
\psfrag{0.003}[Cr][Cr]{\scriptsize{}}
\psfrag{0.005}[Cr][Cr]{\scriptsize{}}
\psfrag{0.007}[Cr][Cr]{\scriptsize{}}
\psfrag{j}[Tc][Tc]{$j$}
\psfrag{I}[Tc][Tc]{$\tilde{I}(j,1000)$}
\psfrag{ji}[Tc][Tc]{}
\psfrag{Ii}[Tc][Tc]{}
\def\stackalignment{r}
\topinset{\includegraphics[width=0.45\textwidth]{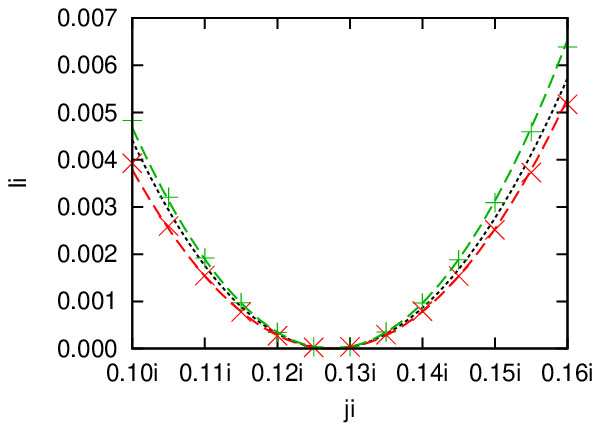}}{\includegraphics[width=0.8\textwidth]{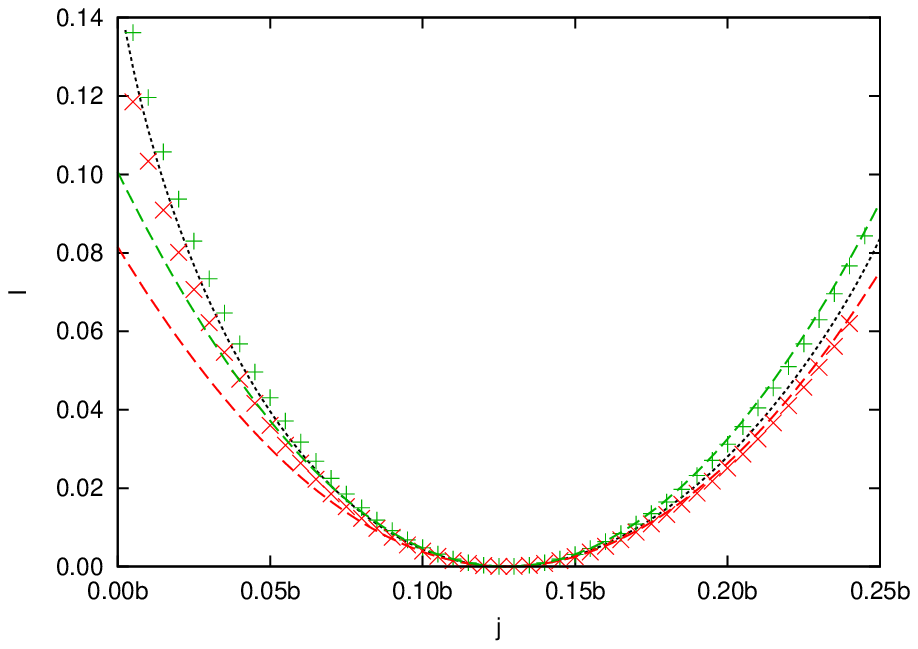}}{0.05\textwidth}{0.15\textwidth}
\caption{Mathematica results for LD-phase $\tilde{I}(j,1000)$ in case $\alpha(j)=\alpha_0 e^{\kappa(j-\bar{j}_{\alpha_0,\beta})}$ with $\alpha_0=0.15$,  $\kappa=0.5$ (red $\times$ symbols) and $\kappa=-0.5$ (green $+$ symbols); initial condition $t_0=1$, $j_0=0$.  Black dotted line is exact expression~\eref{e:TASEPrf} for $I_{\alpha_0}(j)$; coloured dashed lines are Gaussian expansions of $\tilde{I}(j)$ for $\kappa=\pm 0.5$.  Inset shows close-up around mean current.}
\label{f:posneg}
\end{figure}

As a second example, we take a current dependence which illustrates the possibility of negative, as well as positive, feedback.  Specifically we set
\begin{equation}
\alpha(j)=\alpha_0 e^{\kappa(j-\bar{j}_{\alpha_0,\beta})}
\end{equation}
where $\bar{j}_{\alpha_0,\beta}$ is the mean current of a Markovian TASEP with boundary rates $\alpha_0$ and $\beta$.  Note that $\alpha(j)$ thus has a different expression in each of the three regimes of the $(\alpha_0,\beta)$ phase diagram.  For any choice of these boundary rates it is easy to see that $j^*=\bar{j}_{\alpha_0,\beta}$ is a fixed point for all $\kappa$ and, in fact, using the now-established method we find that for $\kappa<8$ this fixed point is always stable.  In other words, for $\kappa<8$ the mean current and phase diagram are identical to the underlying Markovian model but, of course, the fluctuations are different.  In the LD phase, we have
\begin{equation}
A^* = \kappa \alpha_0 (1 - 2 \alpha_0)
\end{equation}
and so, for $\kappa>4$, there is a superdiffusive subregime centred around $\alpha_0=1/4$. 
On the other hand, for $\kappa<4$ we predict diffusive fluctuations throughout the LD phase with modified effective diffusion coefficient
\begin{equation}
\frac{D^*}{1-2A^*} = \frac{\alpha_0(1-\alpha_0)(1-2\alpha_0)}{1-2\alpha_0(1-2\alpha_0)\kappa}.
\end{equation}
In accordance with intuition, negative values of $\kappa$ act to suppress fluctuations and reduce the width of the distribution about the mean current while positive values promote fluctuations and increase the width of the distribution.  This is confirmed by the results shown in figure~\ref{f:posneg} which again demonstrate that the Gaussian approximation agrees closely with the full numerical minimization for small fluctuations but not for large ones (especially below the mean).  

\section{Discussion}
\label{s:dis}

In this paper we have investigated some aspects of a class of interacting particle systems where the rates depend on the time-averaged current $\mathcal{J}/t$.  This memory dependence is effectively a form of feedback which can act to suppress or enhance fluctuations. 
In particular, we here considered the application of a recently proposed ``temporal additivity principle''~\cite{Me09} for obtaining the large deviation rate function for current fluctuations in such non-Markovian models via a minimization involving the Markovian rate function.  Using a heuristic analysis based on fixed points of the dynamics we detailed how a Gaussian approximation for the behaviour of small fluctuations emerges from the full minimization problem and were thus able to highlight the conditions for long-time superdiffusive behaviour.  Whilst this approach fails in general for large fluctuations, it nevertheless provides a means to gain some information about the effects of memory even when the full Markovian rate function is unknown or analytical minimization impossible.   This claim was corroborated by checking the predictions with simulation data for the current mean and variance in a paradigmatic exclusion process model, as well as comparing corresponding exact numerical minimization results.   In order to explore further the underlying assumptions for the temporal additivity principle and its approximation, it would be interesting both to put the central arguments of this paper on a more rigorous mathematical footing and to develop computational methods (perhaps along the lines of the ``cloning'' algorithm~\cite{Giardina06,Lecomte07} for Markovian models) to efficiently access the full rate function in simulations.

Although our chief example here was the \emph{totally} asymmetric simple exclusion process one can also apply similar considerations to models with partially asymmetric dynamics.  Indeed, both the range of applicability of the Gaussian approximation and the stability of numerics would potentially be improved without the cut-off at $j=0$.  In the context of jumps in both forward and backward directions, a topical question is whether one finds a so-called fluctuation relation~\cite{Evans93,Gallavotti95,Lebowitz99} governing the probabilities of positive and negative currents.  Within the Gaussian approximation developed above (for $A^*<1$), we find
\begin{equation}
\fl
\frac{\mathrm{Prob}(\mathcal{J}_t/t=-j)}{\mathrm{Prob}(\mathcal{J}_t/t=j)} \sim 
\cases{
\exp\left[-\frac{2(1-2A^*)j^*}{D^*} \times jt \right] & for $A^*< \frac{1}{2}$ \\
\exp \left[-\frac{2(2A^*-1)j^*}{D^*}t_0^{2A^*-1}\times j t^{2-2A^*}\right]  & for $A^*> \frac{1}{2}$. \\
}
\end{equation}
This suggests the standard Gallavotti-Cohen-type fluctuation symmetry for $A^*< {1}/{2}$ and a modified form for $A^*> {1}/{2}$. The latter is reminiscent of a similar finding for anomalous dynamics in a different setting~\cite{Chechkin09} but a word of caution is necessary here.  Any Gaussian distribution for the current will necessarily have a ratio between positive and negative currents whose exponent is linear in $j$.    This does not guarantee that the same symmetry holds in the non-Gaussian tails of the distribution which are neglected by this approximation.   For time-homogeneous CTRWs or semi-Markov processes (with finite state space and finite mean waiting time), earlier work~\cite{Esposito08,Andrieux08d} asserts that a sufficient condition for the standard symmetry in the full current distribution (arising from a time-reversal relation at the level of microscopic trajectories) is the ``direction-time independence'' property~\cite{Qian06,Maes09b}.  In our framework, an exactly solvable model with the analogous condition that the ratio of jumps left and right is a constant (see also~\ref{A:single}), was indeed found in~\cite{Me09} to obey the symmetry.  
It would be interesting to see if there are similar necessary/sufficient conditions for a modified symmetry relation in the case of superdiffusive fluctuations.  However, in general it is not clear that any such relation exists, much less that it has a meaningful interpretation in terms of entropy (cf.\ the discussion in~\cite{Esposito08}).

Other scenarios worthy of closer attention include fluctuations in models with multiple stable fixed points and fluctuations beyond dynamical phase transitions.  In the latter case, one anticipates the possibility of non-convex rate functions corresponding to non-differentiable points in the scaled cumulant generating function.\footnote{For a mathematical demonstration of a non-convex rate function appearing in another type of non-Markovian model, see~\cite{Duffy08}.}  Usually for Markovian models, a Maxwell-type construction gives phase separation in time and a linear section in the rate function but the introduction of long-range temporal correlations means the phase boundary may acquire a finite probabilistic cost even in the long-time limit.  This is completely analogous to the manner in which long-range spatial correlations can give rise to non-concave entropies in equilibrium~\cite{Campa09}.  Finally we remark that, as already mentioned in~\cite{Me09} the additivity formalism should also be applicable to intrinsically non-Markovian models, such as the Alzheimer random walk~\cite{DaSilva05,Cressoni07,Kenkre07}, but with a non-local minimization problem involving delay differential equations.  The full analytical treatment of such problems appears an even more formidable task but a stability analysis of the dynamics could provide some hope.

Understanding fluctuations in systems with memory is clearly important from both foundational and practical viewpoints but there is much work still to be done in establishing connections between different approaches (especially from complementary mathematics and physics traditions) as well as in forging new ground.   In this context we expect that the workhorses of statistical mechanics such as random walk models and exclusion processes will continue to play an important role. 

\ack

The author is grateful to many colleagues for discussions which have contributed to the development of this material over a number of years. Particular thanks are due to Ajeet Sharma and Debashish Chowdhury for pointing out the connection to~\cite{Sharma11b}, as well as to Hugo Touchette and Massimo Cavallaro for detailed comments on various aspects.  The work has also benefited from the kind hospitality of several research centres especially the Galileo Galilei Institute for Theoretical Physics (GGI) Florence and the National Institute for Theoretical Physics (NITheP) Stellenbosch.

\appendix
\section{Details of a single-particle model}
\label{A:single}

Whilst the bulk of this article is concerned with \emph{many}-particle systems, we here illustrate the formalism by presenting some details of the calculations for a single random walker on an infinite one-dimensional lattice.  We focus in particular on unidirectional dynamics where the particle hops in continuous time to the right only with a rate $v$.   In the memoryless case the rate function for the number of jumps made by the particle is just
\begin{equation}
I_v(j)=v-j+j\ln\frac{j}{v} \quad \textrm{for}~~j\geq0
\end{equation}
which is easily obtained from the limiting behaviour of a Poisson process.  Notice that this is a convex function with a zero at the mean current $\bar{j}_v=v$.   We now modify the picture by considering a rate $v(j)$ with some functional dependence on the time-averaged past current $j$. 

First, in order to understand the correspondence to the CTRW picture we examine the waiting time distribution.  We let $\tau$ be the time of the last jump with a time-averaged current immediately afterwards of $q$ (i.e., the particle last jumped to position $q\tau$) and seek to find the cumulative distribution function (cdf) $F_{q,\tau}(s)$ of the waiting time $s$ until the next jump.  In fact, it turns out to be more convenient to study the complementary cdf $\tilde{F}_{q,\tau}(s)$ which is just the survival function giving the probability that the particle has not jumped up to time $s$.  Now, since the particle's position does not change, the rate at which the next jump takes place depends on time as $v(q\tau/t)$ where $t=\tau+s$.  Hence we trivially have
\begin{equation}
\frac{\tilde{F}_{q,\tau}(s)}{ds}=-v\left(\frac{q\tau}{\tau+s}\right) \tilde{F}_{q,\tau}(s)
\end{equation}
with formal solution
\begin{equation}
\tilde{F}_{q,\tau}(s)=\exp\left\{-\int_0^s v\left(\frac{q\tau}{\tau+u}\right) \, du \right\} \quad \textrm{for}~~s\geq0.  \label{e:formal}
\end{equation}
For specific forms of $v(j)$ one can then calculate the waiting time distribution explicitly.  For example, specialising to the linear form $v(j)=aj+b$ (with $a>0$ and $b>0$) yields
\begin{equation}
\tilde{F}_{q,\tau}(s) =\left( \frac{\tau}{\tau+s} \right)^{aq\tau} e^{-bs} \quad \textrm{for}~~s\geq0 \label{e:wait}
\end{equation}
so that the usual Markovian exponential decay is modified by a power-law prefactor.  Note that, in contrast to a standard (time-homogeneous) CTRW, the waiting time distribution has an explicit dependence on the last jump time $\tau$. 

As an aside, we remark that by an analogous procedure one can calculate the waiting times for a bidirectional random walker with right and left rates given respectively by $v_R(j)$ and $v_L(j)$.  In fact, the distribution of the time between jumps (in any direction) is just given by the expression~\eref{e:formal} with the replacement of the function $v(j)$ by $v_L(j) + v_R(j)$.  A time-inhomogeneous generalization of the so-called ``direction-time independence'' condition~\cite{Qian06,Maes09b} that each transition rate can be written as the product of an individual waiting-time independent probability and a common factor giving the decay of the survival probability, would then seem to require that $v_R(j)/v_L(j)$ is a $j$-independent constant, i.e., that the two rates have the same functional dependence on the current (up to a multiplicative constant).

Returning to the unidirectional case, the Euler-Lagrange equation for the optimal path $q(\tau)$ minimizing the integral~\eref{e:mldp} is
\begin{equation}
v'(q)\left(1-\frac{q}{v}\right) - \frac{2 \tau q'}{q+ \tau q'} - \frac{ \tau^2 q''}{q+\tau q'} = 0.
\end{equation}
In the special case $v(j)=aj$ this is a linear differential equation which is straightforwardly solved~\cite{Me09}.  For $a<1$ one finds
\begin{equation}
\mathrm{Prob}\left(\frac{\mathcal{J}_t}{t}=j\right) \sim e^{-jt_0^at^{1-a}} \qquad j\geq0, \label{e:aq}
\end{equation}
whereas for $a>1$ there is no stationary state and no large deviation principle.  These findings are easily understood within the fixed point analysis of section~\ref{s:fixed}, indeed the mean current condition $v(j)=j$ yields a single fixed point at $j^*=0$ which is stable for $a<1$ and unstable for $a>1$.  The Gaussian expansion is not applicable here since fluctuations below the fixed point (i.e., with $j<0$) are physically impossible.  However, one can argue directly from~\eref{e:aq} that a transition from subdiffusion to superdiffusion occurs when the slope at the fixed point exceeds 1/2.

For the general linear dependence of $v(j)=aj+b$, with $0<a<1$, there is a stable fixed point at $j^*=b/(1-a)$ and one can carry out the Gaussian expansion of section~\ref{s:fixed} where, for this unidirectional random walker, $D^*=j^*$.  Here, there is still a transition at $a=1/2$ but in the ``weaker'' memory phase with $a<1/2$ the $b$ term means that diffusive fluctuations (rather than subdiffusive) are dominant.
The same approach can be carried out with more complicated current dependence as long as there is a single fixed point around which $v(j)$ has a linear dependence on the past current $j$.  For example, with $v(j)=a\sqrt{j}+b$ (and $a>0$, $b>0$) the fixed point is
\begin{equation*}
j^*=\left(\frac{a+\sqrt{a^2+4b}}{2}\right)^2
\end{equation*}
with a slope 
\begin{equation}
A^*=\frac{1}{1+\sqrt{1+4b/a^2}}
\end{equation}
which is constrained to be less than $1/2$ so the fluctuations are always diffusive, albeit with modified diffusion coefficient $D^*/(1-2A^*)$.  A discussion of similar phenomenology in a many-particle model can be found in section~\ref{s:ASEP} of the main text.

\section*{References}
\bibliographystyle{iopart-num}
\bibliography{/home/rosemary/Work/allref}

\end{document}